\documentclass[11pt]{article}
\usepackage{color}
\usepackage{amssymb}
\usepackage{amsmath}
\usepackage{epsfig}

\definecolor{pink}{rgb}{1,.4,.7}
\definecolor{magenta}{rgb}{1,0,1}
\definecolor{violet}{rgb}{.9,.25,.6}
\definecolor{darkolivegreen3}{rgb}{.6,.8,.35}
\definecolor{maroon3}{rgb}{.8,.26,.56}
\definecolor{mediumorchid}{rgb}{.73,.33,.83}
\definecolor{mediumorchid1}{rgb}{1.,.33,.63}
\definecolor{darkgreen}{rgb}{0.1,.6,.13}
\definecolor{lightyellow}{rgb}{1.,1.,.82}
\definecolor{turquoise}{rgb}{.35,.80,.71}
\definecolor{coral}{rgb}{1.,.6,.21}
\definecolor{orangered}{rgb}{1.,.5,0.}
\definecolor{orange}{rgb}{1.,.65,.1}
\definecolor{blue1}{rgb}{.48,.53,1.}
\definecolor{gold}{rgb}{1.,.85,0.}
\definecolor{darkviolet}{rgb}{.54,.04,.84}

\def\be{\begin{equation}}
\def\ee{\end{equation}}
\def\bea{\begin{eqnarray}}
\def\eea{\end{eqnarray}}

\begin{document}
\begin{center}
\Large \bf {Quintessence From a Decaying Dark Matter}\\
\end{center}

\begin{center}
{\it Houri Ziaeepour\\
{Mullard Space Science Laboratory,\\Holmbury St. Mary, Dorking, Surrey
RH5 6NT, UK.\\
Email: {\tt hz@mssl.ucl.ac.uk}}
}
\end{center}

\medskip

\begin {abstract}
It is a known fact that a quintessence model with 
$w_q < -1$ fits the publicly available Super Nova (SN) type Ia data better 
than a model with cosmological constant or $w_q > -1$. Two types of models 
have this property: Scalar fields with unconventional kinetic term and
models with cosmological constant and a slowly decaying Cold Dark Matter
(CDM). In this work we investigate the possibility of replacing the
cosmological constant in the latter models with gradual condensation of a
scalar field produced during the decay of the CDM and present some 
preliminary results. The advantage of this class of models to the ordinary 
quintessence is that the evolution of the dark energy and CDM are 
correlated and cosmological coincidence problem is solved
or at least reduced to the fine tuning of the coupling between decaying CDM
and quintessence field i.e the Hierarchy problem. Here we show that for part 
of the parameter space these models are consistent with present estimation 
of cosmological parameters.
\end {abstract}

\section {Motivations and Model}
The mystery of the Dark Energy / Cosmological Constant persists despite 
great efforts of  particle 
physicists and cosmologists to find a convincing solution. Since the famous 
article by S. Weinberg, it is well known that candidate models should not 
only explain the smallness of the Cosmological Constant, if it is somehow 
related to Quantum Gravity, but also what is called the coincidence problem 
i.e. why does Dark Energy become dominant quite late in the history of the 
Universe? Models inspired by String Theory like 4-form gauge models can 
describe the smallness of the Dark Energy but not the coincidence problem. 
Anthropical models explain the latter problem but it is very difficult to 
find a natural and convincing particle physics model for them. The same 
problem somehow exists for the alternative to a Cosmological Constant i.e. 
for Quintessence Models. Even if tracker solutions make the model not very 
sensitive to the initial conditions, some fine tuning of the slope of the 
potential is necessary. It is also an open question if both 
inflation and quintessence behavior can be explained by the same field and 
if not, what is their relation and which type of particle physics can provide 
both of them specially in a natural way.\\
Here we suggest an alternative to a primordial quintessence field. There are 
at least two motivations 
for the existence of a Decaying Dark Matter (DDM). If R-parity in SUSY 
models is not strictly conserved, the LSP which 
is one of the best candidates of DM can decay to Standard Model 
particles. Violation of this symmetry is one of the many ways 
for providing neutrinos with very small mass and large mixing angle. 
Another motivation is the search for sources of Ultra High Energy Cosmic 
Rays (UHECRs). In this case, DDM 
must be composed of ultra heavy particles with $M_{dm} \sim 10^{22}-10^{25} 
eV$. In a recent work we have shown that the lifetime of 
UHDM (Ultra Heavy Dark Matter) can be relatively short, i.e. $\tau \sim 10 - 
100 \tau_0$ where $\tau_0$ is the age of the Universe {\color{orangered}(astro-ph/0001137)}.\\
If a very small fraction of the mass of primary DDM particles changes to 
a scalar field with proper self interaction potential, the gradual 
condensation of this field at late time behaves like a quintessential 
matter. The advantage of this model to others is that late time yield of 
this type of energy and its correlation with the amount of Dark Matter comes 
up naturally and the dominance of one with respect to the other at each 
epoch is automatically explained.\\
In the present work we only study the plausibility of this model. We postpone 
a more detailed study to elsewhere. The natural 
choice of cosmological parameters for this model is an initial 
${\rho}_{dm} (t_0) \approx {\rho}_{c}(t_0) - {\rho}_{hot}(t_0)$ where 
$t_0$ is an early time in the history of the Universe. For the result 
presented here we consider it to be the time of decoupling of CMB 
photons; ${\rho}_{hot} = {\rho}_{\gamma} + {\rho}_{\nu}$. The remnants of 
the decay else than 
quintessence field mainly consist of very energetic particles which will 
contribute to the yield of Hot Dark Matter. There is strict constraint 
on the amount of the latter from cosmological observations and the model 
must be consistent with observations. However, one 
should not forget that massive particles like proton/ant-proton and even electrons 
become colder with the expansion of the Universe and at some point they 
are not any more considered as hot. In fact it can be shown that the whole 
effect on the equation of state of the Universe is the reduction of the 
effective $w_q$ of the Cosmological Constant or a quintessence matter {\color{orangered} (astro-ph/0002400)}.\\
We summarize our preliminary results in two following figures. The first 
figure shows ${\chi}^2$ of the fit of Quintessence Models on the publicly 
available Super-Novae Ia data. Models with $w_q < -1$ fit the data better 
than $w_q \geq -1$.\\ 
\begin{figure}[h]
\begin{center}
\psfig{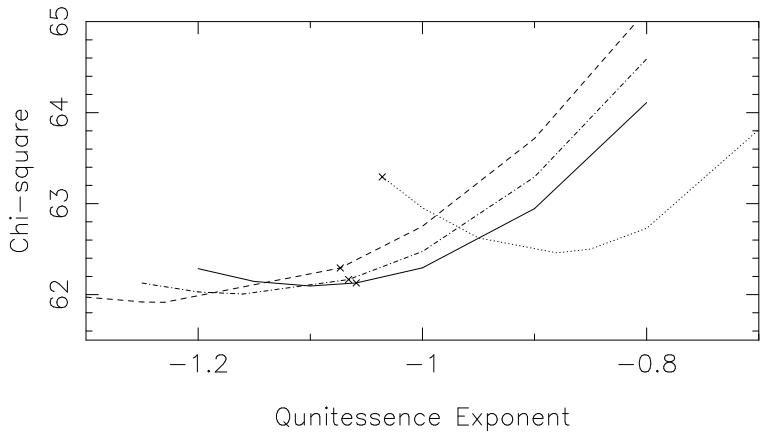}
\caption{$\chi^2$-fit of Quintessence models and equivalent 
Quintessence to a DDM on the Super-Novas Ia data as a function of 
$w_q$: $\Omega_q = 0.67$ (dashed), $\Omega_q = 0.69$ (dash-dot), 
$\Omega_q = 0.71$ (solid) and $\Omega_q = 0.8$ (dotted). The $\chi^2$ of 
equivalent quintessence models to DDMs with lifetime $\tau = 5 \tau_0$ and 
same $\Omega_q$ are shown with crosses. Except $\Omega_q = 0.8$ model, 
other Q-models are very good fit to DDM. For $\Omega_q = 0.8$, a stable 
DM fits the data better, but the fit is poorer than former models. 
\label {fig:fit}}
\end{center}
\end{figure}
The second figure shows the evolution of density of various types of matter 
from decoupling of CMB photons to today for a typical selection of 
parameters. In one hand it shows that it is possible to obtain the present 
``equivalent'' value of cosmological parameters without fine tuning of the 
suggested model. Another conclusion is that the appearance of cosmological 
parameters as measured in the local Universe is very recent i.e. the 
measurement of cosmological parameters at high redshift permits to 
distinguish between this model and other quintessence models.

\section {Discussion and Perspectives}
It is evident that the model presented here can not be believed before 
investigating many 
issues. The first and one of the most important ones is the 
condensation of the scalar field. One should determine the mass and the form 
of the potential and find the region of the parameter space that in a 
natural way 
can lead to a late condensation.  The other issue is that the value of 
$w_q$ for this type of matter can not be constant. This can affect the 
evolution of halos, star formation rate, ionization of IGM etc. and can be 
used to verify the model.
\begin{figure}[h]
\begin{center}
\psfig{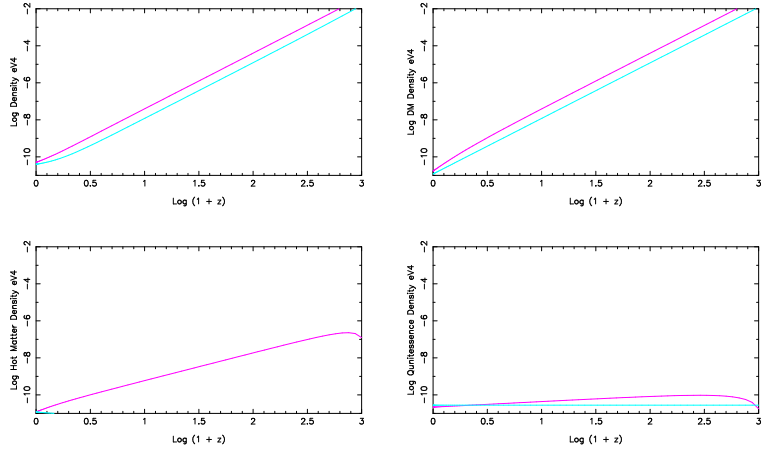}
\caption{Evolution of Total, CDM, Hot and Quintessence matter 
density with redshift (magenta curves). It corresponds to a flat cosmology 
with ${\tau}_{dm} = 1.2 {\tau}_0$ 
where ${\tau}_0$ is the age of the Universe, $w_q = -0.9$ and $f = 10^{-4}$ where 
$f$ is the fraction in energy of the mass of DDM that condensates as a 
quintessence field. For each density the corresponding quantity for a 
``Standard Cosmological'' model with ${\Omega}_{dm} = 0.3$ and 
${\Omega}_{\Lambda} = 0.7$ with only CMB photons and primordial neutrinos as 
Hot Matter is shown (green curves). \label {fig:quin}}
\end{center}
\end{figure}

\end{document}